\documentclass[aps,prl,reprint,superscriptaddress]{revtex4-2}

\usepackage{graphicx} 
\usepackage{amsmath}
\usepackage{amssymb}
\usepackage{tikz}
\usepackage{microtype}
\usetikzlibrary{shapes.geometric}
\newcommand{\figpanel}[3]{\begin{minipage}[t]{#2}\vspace{0pt}\raggedright(#1)\\[1pt]#3\end{minipage}}

\begin{document}

\title{Sample-based Quantum Diagonalization via Density Matrix Renormalization Group}

\author{Yuto Sugimoto}
\affiliation{IBM Research, IBM Japan Ltd., Tokyo 103-8510, Japan}
\affiliation{Department of Physics, Tohoku University, Sendai 980-8578, Japan}
\author{Tomonori Shirakawa}
\affiliation{RIKEN Center for Computational Science (R-CCS), Kobe 650-0047, Japan}
\affiliation{RIKEN Center for Quantum Computing (RQC), Wako 351-0198, Japan}
\affiliation{RIKEN Center for Interdisciplinary Theoretical and Mathematical Science (iTHEMS), Wako 351-0198, Japan}
\affiliation{RIKEN Pioneering Research Institute (PRI), Wako 351-0198, Japan}
\author{Yukio Kawashima}
\affiliation{IBM Research, IBM Japan Ltd., Tokyo 103-8510, Japan}
\author{Toshinari Itoko}
\affiliation{IBM Research, IBM Japan Ltd., Tokyo 103-8510, Japan}
\author{Seiji Yunoki}
\affiliation{RIKEN Pioneering Research Institute (PRI), Wako 351-0198, Japan}
\affiliation{RIKEN Center for Computational Science (R-CCS), Kobe 650-0047, Japan}
\affiliation{RIKEN Center for Quantum Computing (RQC), Wako 351-0198, Japan}
\affiliation{RIKEN Center for Emergent Matter Science (CEMS), Wako 351-0198, Japan}

\begin{abstract}
We introduce SqDMRG, which combines sample-based quantum diagonalization (SQD) with the density matrix renormalization group (DMRG).
SqDMRG expands the sampled determinant space into a product of fragment spaces constructed from the samples, and solves the Hamiltonian in this product space using DMRG.
We demonstrate the method on benzene pentamer and decamer clusters using classical reference samples, and on the pentamer using quantum samples obtained from the IBM Quantum Nighthawk r2 processor \texttt{ibm\_phoenix}.
It achieves sub-kcal/mol accuracy with far fewer configurations than direct diagonalization.
\end{abstract}

\maketitle

\section{Introduction}
Accurate ground-state energy calculations are central to predictive simulations of molecular and material systems, yet their computational cost grows rapidly with system size.
Hybrid quantum-classical algorithms are a promising approach to extending the reach of electronic-structure calculations by combining quantum sampling with classical post-processing.
Among these approaches, sample-based quantum diagonalization (SQD)~\cite{Robledo-Moreno:2024pzz} has emerged as an effective framework for ground-state energy estimation.
SQD augments the original quantum selected configuration interaction method~\cite{Kanno:2023rfr} with configuration recovery to restore the target electron numbers in quantum samples, followed by classical diagonalization of the Hamiltonian in the resulting determinant subspace.
SQD and its variants have enabled calculations on systems beyond the reach of exact diagonalization~\cite{Robledo-Moreno:2024pzz,Shajan:2025umt}.
However, their performance remains strongly dependent on the quality and size of the sampled determinant space, a limitation that becomes particularly severe when the wave function requires a combinatorially large number of determinants.

This challenge is especially pronounced in molecular clusters composed of weakly interacting fragments.
Although the ground state of each fragment may be represented by only a few local determinants, the number of global determinants required to describe the entire cluster grows exponentially with the number of fragments because the full determinant space consists of all combinations of local configurations.
This exponential growth can be avoided by working in products of fragment states, as in the quasi-complete active space approach~\cite{nakanoQuasicompleteActiveSpace2000}, the localized active space self-consistent field (LASSCF) method~\cite{hermesMulticonfigurationalSelfConsistentField2019}, tensor product selected configuration interaction~\cite{abrahamSelectedConfigurationInteraction2020}, and fragment-based variants~\cite{parkerActiveSpaceDecomposition2014,Cheng:2024wui,Larsson:2021rqb} of the density matrix renormalization group (DMRG)~\cite{White:1992zz,2011ARPC...62..465C}.
However, these approaches typically require the fragment basis to be constructed in a separate calculation, for example by solving a fragment Hamiltonian in isolation or within an embedding environment.
This requirement persists when quantum algorithms are used: quantum algorithms may be employed to prepare fragment states~\cite{Otten:2022eya}, while SQD has been used as a fragment solver within LASSCF (LASSQD)~\cite{wangLocalizedSamplebasedQuantum2026} and embedding frameworks~\cite{Shajan:2025umt,Shajan:2024sxw,Merz:2026mml}.
These approaches still rely on separately constructed fragment spaces, and a general framework is lacking for constructing compact fragment spaces directly from determinant samples of the full system while systematically retaining interfragment correlations.

In this work, we introduce sample-based quantum diagonalization via DMRG (SqDMRG), a hybrid quantum-classical method that constructs compact fragment spaces directly from sampled determinants while efficiently capturing correlations between fragments.
Starting from sampled determinants, we extract the configurations occurring on each fragment, construct the product of the resulting fragment spaces, and solve the projected Hamiltonian variationally using DMRG.
In contrast to LASSQD, where the wave function is restricted to a single product of fragment states, SqDMRG incorporates interfragment correlations through the DMRG bond dimension and can be systematically improved by increasing it.
We benchmark SqDMRG using samples drawn from a classical reference wave function for benzene pentamer and benzene decamer clusters and show that it achieves sub-kcal/mol accuracy, an energy error below 1~kcal/mol relative to a near-exact reference energy in the same active space, while requiring far fewer configurations than direct diagonalization.
We further demonstrate SqDMRG with samples measured on the IBM Quantum Nighthawk r2 processor \texttt{ibm\_phoenix} for the benzene pentamer.

\section{Method}
\label{sec:method}

\subsection{Sample-based quantum diagonalization}
We consider a quantum system with a Hamiltonian $\hat H$ acting on a Fock space spanned by Slater determinants $|D\rangle$.
For the molecular systems studied in this work, $\hat H$ is the chemical Hamiltonian in second quantization~\cite{Helgaker:2000qwu},
\begin{equation}
  \hat H = \sum_{pq\sigma} h_{pq}\,a^\dagger_{p\sigma}a_{q\sigma}
  + \frac{1}{2}\sum_{pqrs\sigma\tau} g_{pqrs}\,
    a^\dagger_{p\sigma}a^\dagger_{r\tau}a_{s\tau}a_{q\sigma},
  \label{eq:hamiltonian}
\end{equation}
with one- and two-electron integrals $h_{pq}$, $g_{pqrs}$ over the molecular orbitals $p,q,r,s$ of an active space and spins $\sigma,\tau\in\{\alpha,\beta\}$.

In SQD~\cite{Robledo-Moreno:2024pzz,Kanno:2023rfr}, as in selected configuration interaction (SCI)~\cite{huronIterativePerturbationCalculations1973,harrisonApproximatingFullConfiguration1991,Holmes:2016svu,sharmaSemistochasticHeatBathConfiguration2017}, $\hat H$ is diagonalized in a subspace of selected determinants, but these determinants are sampled from a quantum state.
Sampled $k$ determinants $\mathcal{S}=\{|D_i\rangle\}_{i=1}^{k}$ define the projector $P_\mathcal{S}=\sum_{D\in\mathcal{S}}|D\rangle\langle D|$ onto the selected determinants.
Diagonalizing the projected Hamiltonian in the subspace spanned by these $k$ determinants,
\begin{equation}
  P_\mathcal{S}\hat H P_\mathcal{S}\,|\Psi_\mathcal{S}\rangle = E_\mathcal{S}\,|\Psi_\mathcal{S}\rangle,
  \label{eq:sqd}
\end{equation}
yields an approximate ground-state energy and wave function, with $|\Psi_\mathcal{S}\rangle$ a linear combination of the sampled determinants.
The accuracy of this approximation is controlled by the quality and size of $\mathcal{S}$.
If the exact ground state is sparse in the determinant basis, a small sample can capture its dominant subspace, whereas a dense ground state forces $k$ to grow.
The quality of $\mathcal{S}$ depends on how it is generated.
On a quantum computer, $\mathcal{S}$ is generated by repeated measurement of a circuit that prepares an approximate ground state~\cite{Robledo-Moreno:2024pzz,Kanno:2023rfr}.
We also consider an idealized classical sample, drawn from a converged classical wave function.

A cluster of molecules, in which intramolecular correlation is strong but intermolecular correlation is weak, is an extreme example in which $k$ must grow.
In this case, the ground state of each molecule $A$ may be sparse in the determinant basis, so that only $k_A$ local determinants are needed for a good approximation to that molecule.
Over the global determinant space, however, the number of determinants required for a comparably good approximation grows as $k_A^N$, where $N$ is the number of molecules.
This exponential growth in the required sample size makes direct SCI and SQD impractical for large clusters.

\begin{figure*}[tbp]
\centering
\begin{tikzpicture}[x=1cm, y=1cm,
  op/.style={draw, thick, circle, fill=cyan!35, minimum size=4mm,
             inner sep=0pt},
  pr/.style={draw, thick, isosceles triangle, isosceles triangle stretches,
             fill=cyan!15, minimum width=2.05cm, minimum height=5mm,
             inner sep=0pt},
  cop/.style={draw, thick, circle, fill=teal!40, minimum size=7mm,
              inner sep=0pt},
  lab/.style={font=\scriptsize},
  l/.style={thick}
]
  \node[font=\footnotesize] at (-1.7,0) {$P_\otimes\hat H P_\otimes=$};
  \draw[l] (-0.45,0) -- (6.5,0);
  \foreach \i in {0,...,11} {
    \draw[l] (0.55*\i,-0.72) -- (0.55*\i,0.72);
    \node[op] at (0.55*\i,0) {};
  }
  \node[lab, left] at (0,0.5) {};
  \foreach \f in {1,2,3} {
    \pgfmathsetmacro{\xc}{0.55*(4*(\f-1)+1.5)}
    \draw[l] (\xc,1.1) -- (\xc,1.6);
    \draw[l] (\xc,-1.1) -- (\xc,-1.6);
    \node[pr, shape border rotate=90,  anchor=south] at (\xc, 0.72) {};
    \node[pr, shape border rotate=270, anchor=north] at (\xc,-0.72) {};
    \node[font=\footnotesize, right] at (\xc+0.25, 1.35) {$P_\f$};
    \node[font=\footnotesize, right] at (\xc+0.25,-1.35) {$P_\f$};
    \node[lab, above] at (\xc,1.6) {$n_\f$};
  }
  \foreach \x in {1.925,4.125} {\draw[dashed] (\x,-1.4) -- (\x,1.4);}
  \node[font=\footnotesize] at (7.3,0) {$=$};
  \draw[l] (7.8,0) -- (10.9,0);
  \foreach \f/\x in {1/8.1, 2/9.35, 3/10.6} {
    \draw[l] (\x,-0.85) -- (\x,0.85);
    \node[cop] at (\x,0) {};
    \node[lab, above] at (\x,0.85) {$n_\f$};
  }
\end{tikzpicture}
\caption{The figure shows the Hamiltonian projected by $P_\otimes$, the projector onto the product of the sampled local determinant sets, for the example of Eq.~\eqref{eq:split}.
On the left-hand side, $\hat H$ is written as a matrix product operator with one site per spatial orbital, and each orbital has a physical leg of dimension 4.
The dashed lines mark the fragment boundaries.
For each fragment $A$, the local projector $P_A$ projects onto $\mathcal{S}_A$, the set of $n_A$ unique local determinants sampled for that fragment.
It acts from both sides on the four physical legs.
On the right-hand side, contracting each fragment gives an $N$-site matrix product operator whose physical legs have dimension $n_A$.
When $n_A$ is large, this local basis is truncated using eigenvectors with the largest eigenvalues of the fragment reduced density matrix.}
\label{fig:schematic}
\end{figure*}
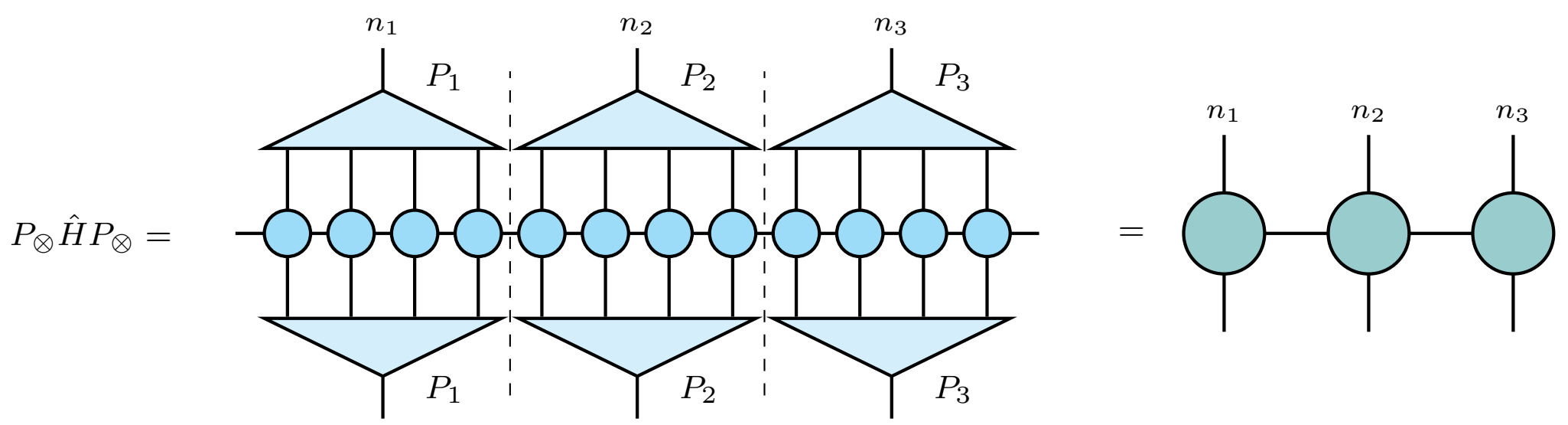

\subsection{SqDMRG}
To reduce the required sample size, we consider effectively increasing the number of samples by splitting the sampled determinants into fragments.
For a cluster of molecules, we partition the orbitals into $N$ fragments, one for each molecule, and write each determinant as $|D\rangle=\bigotimes_{A=1}^{N}|D_A\rangle$ with a local determinant $|D_A\rangle$ of fragment $A$.
In Born sampling of a state of the whole system, the weight is spread over many global determinants, each of which is observed with a small probability, so that $\mathcal{S}$ may miss most of the $k_A^N$ required combinations.
In contrast, the probability of observing a local determinant $|D_A\rangle$ is the sum of the weights of all its combinations with the configurations of the other fragments, which is the diagonal element of the reduced density matrix of this state on fragment $A$.
The sample $\mathcal{S}$ is therefore likely to contain the local determinants important for each fragment in the sampled state.
We can thus expect that taking every combination of the local determinants that appear in $\mathcal{S}$ supplies the combinations missing from $\mathcal{S}$.
To illustrate this, we consider a simple example with $N=3$ fragments of four orbitals.
Labeling the occupation of each spatial orbital by $0\leftrightarrow|0\rangle$, $1\leftrightarrow|\alpha\rangle$, $2\leftrightarrow|\beta\rangle$, and $3\leftrightarrow|\alpha\beta\rangle$, we suppose that the exact ground state is
\begin{align*}
  |\Psi\rangle
  ={}& c_1|3300\,3300\,3300\rangle + c_2|3120\,3300\,3300\rangle \\
  &+ c_3|3300\,3030\,3300\rangle + c_4|3300\,3300\,3210\rangle \\
  &+ c_5|3120\,3030\,3210\rangle,
\end{align*}
and that we sample it twice and obtain
\begin{equation*}
  \mathcal{S}
  =
  \left(\;
  \smash[b]{\underbrace{\begin{array}{c}3300\\3120\end{array}}_{\text{fragment }1}}
  \quad
  \smash[b]{\underbrace{\begin{array}{c}3300\\3030\end{array}}_{\text{fragment }2}}
  \quad
  \smash[b]{\underbrace{\begin{array}{c}3300\\3210\end{array}}_{\text{fragment }3}}
  \;\right)
  \vphantom{\underbrace{\begin{array}{c}3300\\3120\end{array}}_{\text{fragment }1}},
\end{equation*}
where each row of $\mathcal{S}$ is a sampled determinant and the braces mark the fragment boundaries.
Three of the five determinants in $|\Psi\rangle$ thus do not appear in $\mathcal{S}$, although each of their local determinants does.
We can therefore recover them by collecting the unique local determinants of each fragment and taking their tensor product,
\begin{equation}
  \mathcal{S}_1\otimes\mathcal{S}_2\otimes\mathcal{S}_3
  =
  \begin{pmatrix}3300\\3120\end{pmatrix}
  \otimes
  \begin{pmatrix}3300\\3030\end{pmatrix}
  \otimes
  \begin{pmatrix}3300\\3210\end{pmatrix},
  \label{eq:split}
\end{equation}
where $\mathcal{S}_1$, $\mathcal{S}_2$, and $\mathcal{S}_3$ are the sets of unique local determinants of fragments 1, 2, and 3 that appear in $\mathcal{S}$.
This product set contains $2^3=8$ determinants, including all the five determinants in $|\Psi\rangle$.
In general, we collect the $n_A$ unique local determinants of fragment $A$ that appear in $\mathcal{S}$ into a set $\mathcal{S}_A$, and the product set $\mathcal{S}_1\otimes\cdots\otimes\mathcal{S}_N$ includes every combination of the sampled local determinants, so that it contains $\mathcal{S}$ and clearly spans a larger space than $\mathcal{S}$.
The effective sample size thus increases from $k$ to $\prod_A n_A$.
However, when $N$ is large, this enlarged set is too large for direct diagonalization as in Eq.~\eqref{eq:sqd}, and the problem is then how to diagonalize $\hat H$ in it.

\begin{figure*}[tbp]
\centering
\figpanel{a}{0.55\textwidth}{\includegraphics[width=\linewidth]{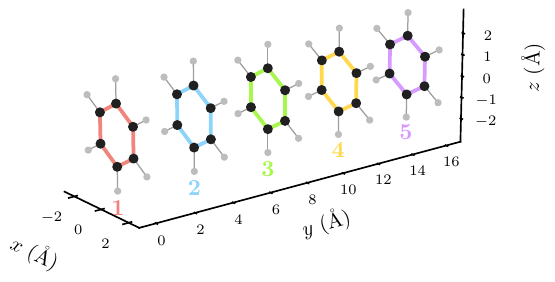}}\hspace{0.06\textwidth}
\figpanel{b}{0.32\textwidth}{\includegraphics[width=\linewidth]{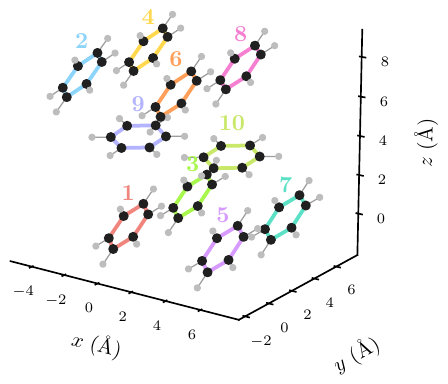}}
\caption{Geometries of (a) the benzene pentamer and (b) the benzene decamer.
Each benzene molecule forms one fragment, and the numbers label the fragments.}
\label{fig:systems}
\end{figure*}

\begin{figure*}[tbp]
\centering
\figpanel{a}{0.26\textwidth}{\includegraphics[width=\linewidth]{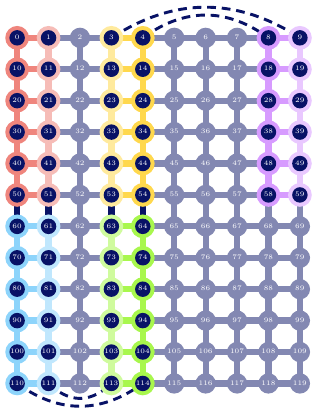}}\hspace{0.03\textwidth}%
\figpanel{b}{0.66\textwidth}{\includegraphics[width=\linewidth]{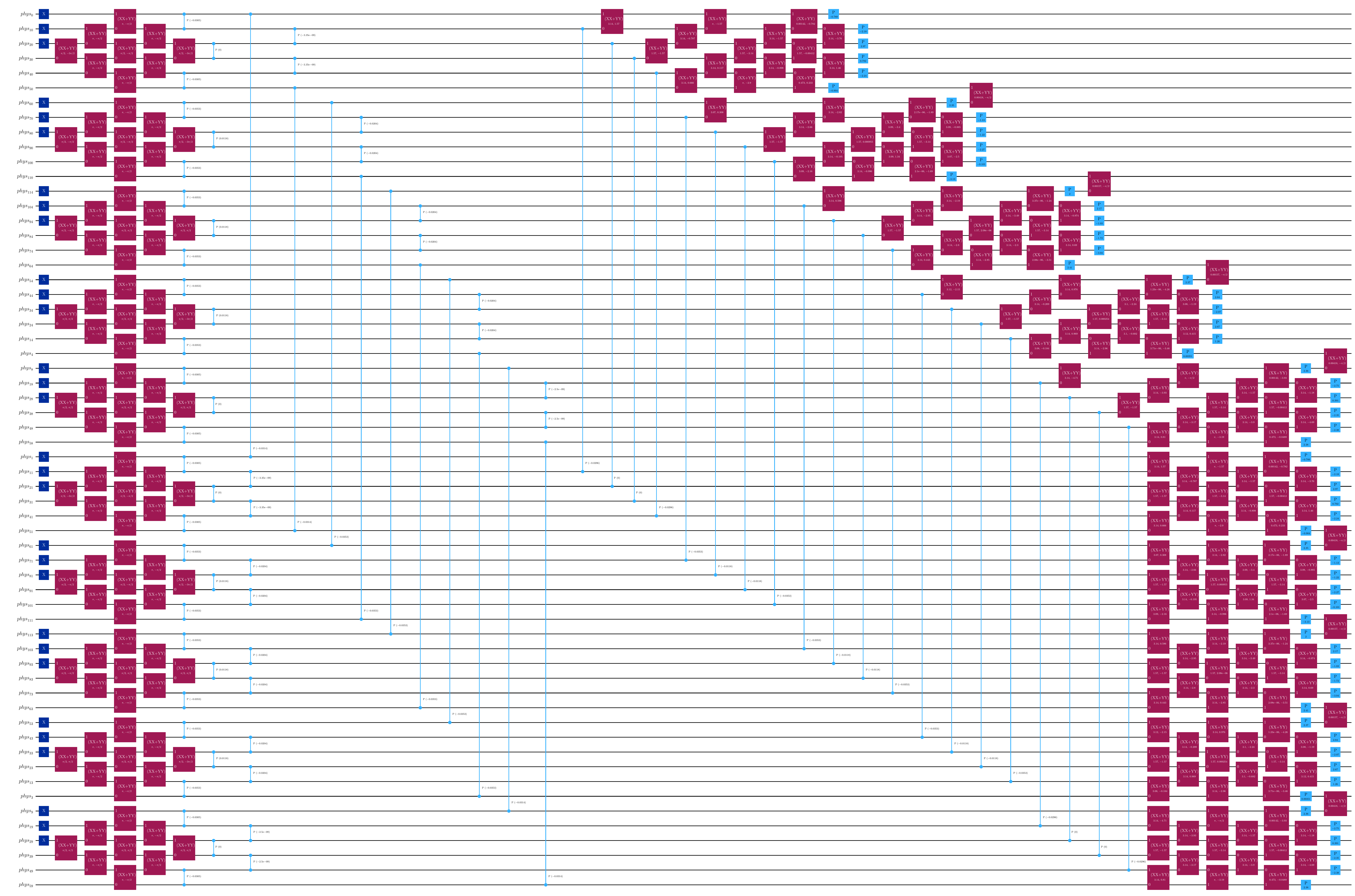}}
\caption{Quantum sample for the benzene pentamer.
(a) Qubit assignment on the $12\times10$ square lattice of \texttt{ibm\_phoenix}.
The colored qubits and couplers belong to the fragments with the colors of Fig.~\ref{fig:systems}(a), where the two columns of each block hold the $\alpha$ (darker) and $\beta$ (lighter) spin orbitals.
The dark couplers connect neighboring fragments, and the dashed lines connect boundary qubits of neighboring fragments that are not adjacent on the lattice.
The numbers are the physical qubit indices.
(b) Local unitary cluster Jastrow (LUCJ) circuit before transpilation, where each wire is labeled by the physical qubit in (a).}
\label{fig:circuit}
\end{figure*}

\begin{figure}[tbp]
\centering
\includegraphics[width=0.8\linewidth]{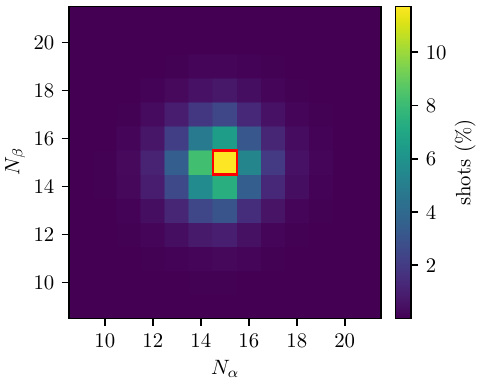}
\caption{Distribution of the electron numbers $(N_\alpha,N_\beta)$ in the $10^7$ raw shots from \texttt{ibm\_phoenix} before configuration recovery.
The red square marks the target sector $N_\alpha=N_\beta=15$.}
\label{fig:raw}
\end{figure}

\begin{figure*}[tbp]
\centering
\includegraphics[width=0.9\textwidth]{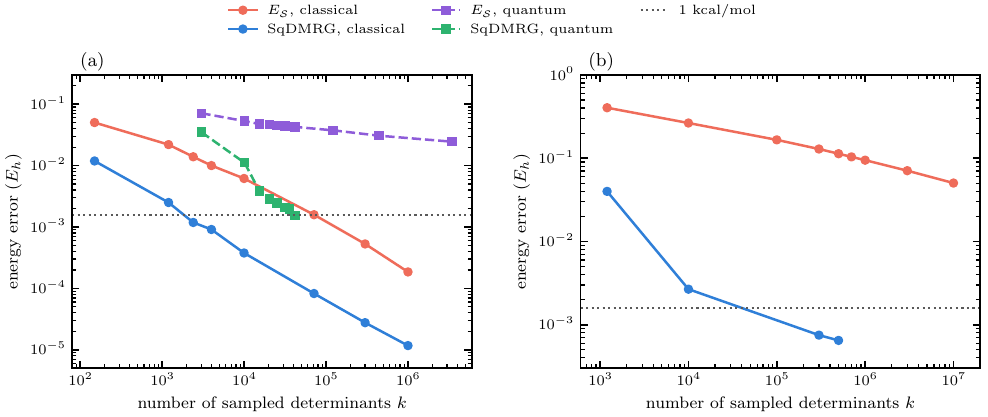}
\caption{Energy error relative to the reference DMRG energy as a function of the number of sampled determinants $k$ for (a) the benzene pentamer and (b) the benzene decamer.
The energy $E_\mathcal{S}$ is obtained by diagonalizing $\hat H$ in $\mathcal{S}$, and SqDMRG uses the same $\mathcal{S}$.
Circles and squares show the classical and quantum samples, and the dotted line at 1 kcal/mol marks the threshold for sub-kcal/mol accuracy.}
\label{fig:ladder}
\end{figure*}

However, this enlarged sample set is not an arbitrary set of determinants but the tensor product of the local sets $\mathcal{S}_A$, which has the same structure as the Hilbert space of an $N$-site lattice with local dimensions $n_A$.
A Hilbert space with such a structure is what DMRG handles efficiently, because DMRG optimizes the ground state site by site without enumerating all the $\prod_A n_A$ states.
This suggests that, by treating the local set $\mathcal{S}_A$ of each fragment as one site and performing DMRG~\cite{parkerActiveSpaceDecomposition2014,Cheng:2024wui,Larsson:2021rqb,White:1999thl,nishioRankoneBasisMade2019,larssonMinimalMatrixProduct2020}, we can obtain the same ground state as the diagonalization in the product set, if the bond dimension is large enough.
In fact, the idea of restricting the Hilbert space of a large site to a given set of determinants has already been mentioned in large-site DMRG~\cite{Larsson:2021rqb,zhaiBlock2ComprehensiveOpen2023}.
We follow this idea, so that the local basis of each site is taken directly from the sample.
To realize this restriction, we introduce the local projector $P_A=\sum_{D_A\in\mathcal{S}_A}|D_A\rangle\langle D_A|$ onto the sampled determinants of fragment $A$.
The tensor product of these local projectors gives the projector onto the product set,
\begin{equation}
  P_\otimes=P_1\otimes P_2\otimes\cdots\otimes P_N.
  \label{eq:Pprod}
\end{equation}
Since the product set contains $\mathcal{S}$, the variational principle ensures that the exact ground-state energy of $P_\otimes\hat H P_\otimes$ is lower than or equal to $E_\mathcal{S}$ of Eq.~\eqref{eq:sqd}.
SqDMRG can therefore be understood as an improved solver for SCI and SQD with a given sample.

In practice, we write the projected Hamiltonian $P_\otimes\hat H P_\otimes$ as an $N$-site matrix product operator (MPO), or equivalently, directly multiply the MPO of the full Hamiltonian $\hat H$ by $P_\otimes$ from both sides, as illustrated in Fig.~\ref{fig:schematic}.
We then optimize a matrix product state (MPS) on this $N$-site lattice by DMRG.
The MPS determines the coefficients of the determinants in the product set, and its bond dimension $M$ represents the correlation between fragments.
In fact, when the interaction between fragments is weak, as in the clusters considered here, ASD-DMRG and BIPS-DMRG~\cite{parkerActiveSpaceDecomposition2014,Cheng:2024wui} have shown that this grouping keeps the strong intrafragment correlation within each site and leaves only the weak interfragment correlation to the MPS, so that only a small $M$ is needed.
Even with a small $M$, however, such cluster DMRG calculations are in general expensive because the local dimension $n_A$ of each site is large, and we therefore use one-site DMRG as in these methods~\cite{whiteDensityMatrixRenormalization2005,hubigStrictlySinglesiteDMRG2015}, which is cheaper than two-site DMRG for a large local dimension.

Even with one-site DMRG, the calculation becomes expensive when a fragment is large or has many unique local determinants in $\mathcal{S}$, because $n_A$ can then become large.
For such fragments, we reduce $n_A$ by diagonalizing $\hat H$ in $\mathcal{S}$ as in Eq.~\eqref{eq:sqd}, computing the reduced density matrix $\rho_A$ of each fragment from its ground state $|\Psi_\mathcal{S}\rangle$, and keeping the $m_A$ eigenvectors of $\rho_A$ with the largest eigenvalues as the new local basis of fragment $A$~\cite{Cheng:2024wui,White:1992zz,White:1999thl,zhangDensityMatrixApproach1998}.
Since $\rho_A$ acts only on the local determinants in $\mathcal{S}_A$, the retained eigenvectors are linear combinations of them, and their products remain in the product set.
The local basis is thus determined by the sample alone, without a separate calculation on each fragment as in previous methods~\cite{hermesMulticonfigurationalSelfConsistentField2019,abrahamSelectedConfigurationInteraction2020,parkerActiveSpaceDecomposition2014,Cheng:2024wui,Otten:2022eya,wangLocalizedSamplebasedQuantum2026}.

SqDMRG thus involves three approximations to the exact ground state.
First, the product set may not contain all determinants that contribute to the exact ground state.
Second, the truncation to $m_A$ local states restricts the wave function to a subspace of the product set.
Third, the finite bond dimension $M$ limits the entanglement between fragments that the MPS can represent.
The bound by $E_\mathcal{S}$ therefore holds strictly only without the truncation and in the limit of large $M$, although SqDMRG numerically gives energies below $E_\mathcal{S}$ in all our calculations.
We remark that, because of these approximations, SqDMRG cannot be more accurate than converged DMRG on the full active space.

\section{Numerical results}
We apply SqDMRG to clusters of benzene molecules, which are an example of the clustered systems discussed in the Introduction.
We consider a stacked benzene pentamer ($N=5$) and a benzene decamer ($N=10$), shown in Fig.~\ref{fig:systems}, in which each benzene molecule forms one fragment with six $\pi$ orbitals and six electrons.
The active spaces are therefore CAS(30e,30o) and CAS(60e,60o).
In the pentamer, the five molecules are stacked linearly with an intermolecular distance of 4.0~\AA, and each monomer geometry is optimized at the B3LYP/def2-TZVPP level~\cite{beckeDensityfunctionalThermochemistryIII1993,Lee:1988zzc,Weigend:2005akm} with PySCF~\cite{Sun:2018lpq,Sun:2020jul}.
We compute the Hartree--Fock orbitals with the def2-SVP basis set and localize them onto the fragments with the Pipek--Mezey method~\cite{pipekFastIntrinsicLocalization1989} in BAGEL~\cite{shiozaki2018bagel}, and select the six $\pi$ orbitals of each fragment.
In the decamer, eight molecules are located at the vertices of a parallelepiped, and the other two are located at the centers of a pair of opposite faces.
We take the geometry and the integrals from the FermiCG data set~\cite{FermiCG}, where the six $\pi$ orbitals of each fragment are obtained by projecting the Hartree--Fock orbitals with the STO-3G basis set~\cite{hehre1969self} onto the fragment.
To obtain the reference energies, we perform DMRG calculations on the full active space with block2~\cite{zhaiBlock2ComprehensiveOpen2023} and converge them to $10^{-8}\,E_\mathrm{h}$ for the pentamer and to about $10^{-5}\,E_\mathrm{h}$ for the decamer.
These convergence errors are sufficiently smaller than the energy errors discussed below.

To benchmark SqDMRG, we use two types of samples.
One is the idealized classical sample introduced in the Method section, which we use for both the pentamer and the decamer.
The other is a sample generated on quantum hardware, which we use only for the pentamer.
For each sample $\mathcal{S}$, we compare the energy $E_\mathcal{S}$ obtained by diagonalizing $\hat H$ in $\mathcal{S}$ as in Eq.~\eqref{eq:sqd} with the energy of SqDMRG on the same $\mathcal{S}$.

The classical sample is taken from the reference DMRG wave function.
We expand this wave function as $\sum_D c_D|D\rangle$ and take the $k$ determinants with the largest weights $|c_D|^2$ as $\mathcal{S}$.
This set corresponds to the $k$ most frequent determinants in Born sampling of the ground state in the limit of infinitely many samples, and therefore contains no sampling noise.

For the quantum sample, we use a local unitary cluster Jastrow (LUCJ) circuit~\cite{mottaBridgingPhysicalIntuition2023} applied to the Hartree--Fock determinant.
Following Ref.~\cite{Robledo-Moreno:2024pzz}, we obtain its parameters from coupled-cluster singles and doubles (CCSD)~\cite{Purvis:1982ueh}, which we solve for the full active space with PySCF~\cite{Sun:2018lpq,Sun:2020jul} and transform to the localized orbitals.
For each fragment, we keep only the amplitudes whose occupied and virtual orbitals all belong to the fragment.
The double factorization of these amplitudes~\cite{mottaLowRankRepresentations2021} gives a single-layer LUCJ circuit, which consists of an orbital rotation, a Jastrow factor, and a final orbital rotation within the fragment.
In this Jastrow factor, we keep only the same-spin interactions between neighboring orbitals and the opposite-spin interactions within each orbital.
Because these fragment circuits do not move electrons between fragments, we add to the final orbital rotation a Givens rotation with angle $\kappa_A$ between the last orbital of fragment $A$ and the first orbital of fragment $A+1$.
We set $\kappa_A=\arcsin r_A$, where $r_A$ is the ratio of the root-mean-square exchange integral $g_{pqqp}$ between the two fragments to the average of the root-mean-square exchange integrals within them.
The resulting circuit acts on 60 qubits for the 30 spatial orbitals and two spins.
We construct it with ffsim~\cite{sungFfsimFasterSimulation2026} and Qiskit~\cite{Javadi-Abhari:2024kbf} and run it on \texttt{ibm\_phoenix}, an IBM Quantum Nighthawk r2 processor with 120 qubits, as shown in Fig.~\ref{fig:circuit}.
Each fragment occupies a $6\times2$ block of qubits, in which the $\alpha$ and $\beta$ qubits of the same orbital are neighbors, so that the intrafragment interactions act on neighboring qubits.
We take $10^7$ shots, and 11.7\% of them have the correct electron numbers $N_\alpha=N_\beta=15$, as shown in Fig.~\ref{fig:raw}.
Because the fragment-wise LUCJ circuit is shallow, these shots already have a meaningful distribution before configuration recovery.
The Hartree--Fock determinant is the most frequent among them, and the number of shots decreases with the excitation rank from it.
The other shots have wrong electron numbers because of hardware noise, and we restore their electron numbers by configuration recovery~\cite{Robledo-Moreno:2024pzz} with the reference Hartree--Fock occupations.
We then take the $k$ most frequent determinants after a single pass of the recovery as $\mathcal{S}$, without iterating the configuration recovery.

We perform the SqDMRG calculations with ITensor~\cite{Fishman:2020gel} and use $M=64$ for both clusters, $m_A=64$ for the pentamer, and no truncation of the local bases for the decamer.
Because the number of orbitals is large, the $N$-site MPO has a large bond dimension, and we compress it before the DMRG calculation.
Even with this compression, our current implementation is less efficient than conventional DMRG on the full active space.

Figure~\ref{fig:ladder}(a) shows the energy errors for the pentamer.
For the classical sample, SqDMRG gives lower errors than $E_\mathcal{S}$ at every $k$ and reaches sub-kcal/mol accuracy with about 30 times fewer determinants than direct diagonalization.

In Fig.~\ref{fig:ladder}(b), the difference between the two methods becomes larger for the decamer.
The error of $E_\mathcal{S}$ decreases slowly with $k$ and stays far above sub-kcal/mol accuracy even at $k=10^7$.
In contrast, SqDMRG reaches sub-kcal/mol accuracy by $k=3\times10^5$.
This larger improvement for $N=10$ is consistent with the discussion in the Method section, because the number of combinations of local determinants in the product set grows exponentially with $N$.

For the quantum sample of the pentamer in Fig.~\ref{fig:ladder}(a), both methods need more determinants than for the classical sample, because the recovered sample deviates from the ideal classical sample.
Nevertheless, SqDMRG reduces the error monotonically with $k$ and reaches sub-kcal/mol accuracy at $k\approx4\times10^4$, whereas $E_\mathcal{S}$ for the same samples remains far above sub-kcal/mol accuracy up to $k\approx3.4\times10^6$.

\section{Summary}
We have introduced SqDMRG, which constructs a tensor product of fragment spaces directly from sampled determinants and uses DMRG to variationally approximate the ground state of the Hamiltonian projected into this space.
The fragment spaces require no separate fragment calculations, and interfragment correlations are represented by the MPS bond dimension.
In benchmarks using idealized classical samples, SqDMRG reaches an energy error below 1 kcal/mol relative to the reference DMRG energy with about 30 times fewer selected determinants than direct diagonalization for the benzene pentamer.
For the decamer, it reaches this accuracy by $k=3\times10^5$, whereas direct diagonalization remains above the threshold even at $k=10^7$.
Using quantum samples from \texttt{ibm\_phoenix}, SqDMRG reaches the same accuracy for the pentamer with about $4\times10^4$ selected determinants after a single pass of configuration recovery.
It is worth extending the quantum sampling to the decamer and to larger systems, which we leave for future work.
Another future direction is to apply SqDMRG to a single molecule divided into fragments, as in LASSQD~\cite{wangLocalizedSamplebasedQuantum2026}.
We also plan to apply SqDMRG to systems with large fragments, for which conventional DMRG on the full active space also becomes difficult.
For such systems, a comparison with other fragment-based methods~\cite{abrahamSelectedConfigurationInteraction2020,Cheng:2024wui} will further clarify the advantages of SqDMRG.

\begin{acknowledgments}
Y.S. acknowledges support from the Graduate Program on Physics for the Universe (GP-PU) at Tohoku University and from JSPS KAKENHI Grant No. 25KJ0537.
\end{acknowledgments}

\bibliography{refs}

\end{document}